\documentclass[aps,prb,reprint,twocolumn]{revtex4-1}
\usepackage{color,graphicx,calc,url}
\usepackage{dcolumn}
\usepackage{bm}
\usepackage{amssymb}
\usepackage{amsmath}
\usepackage{wasysym}
\usepackage{mathrsfs} 
\usepackage{array}
\usepackage{commath}
\usepackage{float}
\usepackage{textcomp}
\usepackage{braket}
\usepackage[colorlinks=true,
            linkcolor=red,
            urlcolor=blue,
            citecolor=blue]{hyperref}
\usepackage{braket}

\begin{document}
\title{Theoretical studies of enhanced anomalous Nernst effect in Fe$_3$Ga}

\author{Ond\v{r}ej Stejskal}
\email[]{stejskal@karlov.mff.cuni.cz}
\affiliation{Faculty of Mathematics and Physics, Charles University, Prague, Czech Republic}

\author{Martin Veis}
\affiliation{Faculty of Mathematics and Physics, Charles University, Prague, Czech Republic}

\author{Jaroslav Hamrle}
\affiliation{Faculty of Mathematics and Physics, Charles University, Prague, Czech Republic}

\date{\today}

%%% ABSTRACT %%%
\begin{abstract}

The anomalous Nernst effect (ANE) is a member of the extensive family of topological effects in solid state physics. It converts a heat current into electric voltage and originates from the Berry curvature of electronic bands near the Fermi level. Recent results established the Fe$_3$Ga alloy as one of the most promising candidates for applications, due to its flat band structure consisting of rich web of nodal lines. In this theoretical work, we study the effect of deformation of Fe$_3$Ga on the anomalous Nernst effect, which naturally occurs in thin films. Furthermore, we demonstrate that doping, which effectively shifts the position of the Fermi level, can also significantly modify the strength of the effect. Lastly, we provide detailed analysis of the origin of ANE in the electronic structure of Fe$_3$Ga which yields a deeper insight into the generating mechanisms, understanding of which can lead to substantial enhancement of the effect in the future.

%It has been observed that flat band structure consisting of rich web of nodal lines increases the strength of the effect significantly. However, the description of the exact mechanism is currently missing. Here, we provide a detailed analysis of the origin of ANE in the electronic structure of Fe$_3$Ga that goes beyond the common argument of nodality. Utilizing the full vectorial nature of the Berry curvature tensor, we present an analytical expression to describe and estimate the strength of the effect. The detailed understanding of the origin of ANE yields a deeper insight into the generating mechanisms and can lead to substantial enhancement of the effect via band structure engineering methods.
\end{abstract}

\maketitle

\section{Introduction}

In recent years, topological effects have attracted a lot of attention in the solid state research, due to their unique and interesting manifestations with an enormous potential in applications. In magnetic materials, broken time-reversal symmetry induces anomalous linear response effects such as the anomalous Hall effect~\cite{Yao2004,Nagaosa2010,Xiao2010,Ernst2019,Helman2021}, anomalous Nernst effect (ANE)~\cite{Xiao2006,Sakai2018} and the magneto-optic Kerr effect~\cite{Kerr1877,Silber2019}, all being governed by the presence of the Berry curvature of the electronic states.

ANE manifests itself as thermoelectricity, i.e. lossless conversion of heat flow into electricity, which plays a key role in developing novel energy harvesting technologies. The transverse geometry of the effect in ferromagnets offers many advantages to the conventional longitudinal Seebeck effect~\cite{Snyder2008,Hu2019}, but is significantly smaller in magnitude. Currently, huge effort is devoted into searching and identifying materials providing large ANE at zero field.

It has been discovered that ANE originates from the Berry curvature of the conduction electrons at the Fermi level~\cite{Nagaosa2010} and is strongly enhanced with presence of nodal lines and planes for bands at the Fermi level which are then split by the spin-orbit interaction (SOI) to provide large Berry curvatures~\cite{Ikhlas2017,Liu2018_2,Noky2019,Guin2019,Minami2020,Noky2020,Sakai2020,Chen2022,Nakatsuji2022}. Therefore, an extensive amount of high-throughput first-principle calculations were performed in order to find materials with large nodal structures and high density of states at the Fermi energy.

The research has shifted towards ferromagnetic Heuslers~\cite{Reichlova2018,Noky2020}, Fe$_3$X alloys~\cite{Sakai2020,Hamada2021,Chen2022} lately, with Fe$_3$Ga coming out on top~\cite{Nakayama2019,Sakai2020} due to its rich flat band strucure at Fermi energy in the vicinity of point L. This material has been calculated before~\cite{Paduani2011,Sakai2020} but  limited only to stoichiometric undeformed cubic structure.

In this work, we investigate how modifications of the crystal structure of Fe$_3$Ga by both deformation and doping affect the value of ANE. We show that compressive strain is to be sought for in thin film applications.

Furthermore, the electronic structure of DO$_3$ ordered Fe$_3$Ga is thoroughly studied in order to identify the underlying origin of its large ANE. Two distinct sources of the effect are identified in the Brillouin zone and analyzed separately.

%In this work, we investigate the electronic structure of DO$_3$ ordered Fe$_3$Ga in order to identify the underlying origin of its large ANE. This analysis is unique as it aims for deeper understanding of the generating mechanism of the anomalous effects utilizing the full vectorial nature of the Berry curvature tensor. Two distinct sources of the effect are identified in the Brillouin zone and analyzed separately. The topology manifests itself in the proposed analytical formula for estimation of the strength of the effect in certain cases.

%Furthermore, we investigate how modifications of the crystal structure of Fe$_3$Ga by both deformation and doping affect the value of ANE.

\section{Electronic structure of F\lowercase{e}$_3$G\lowercase{a}}

\begin{figure}
\begin{center}
\includegraphics[width=0.5\textwidth]{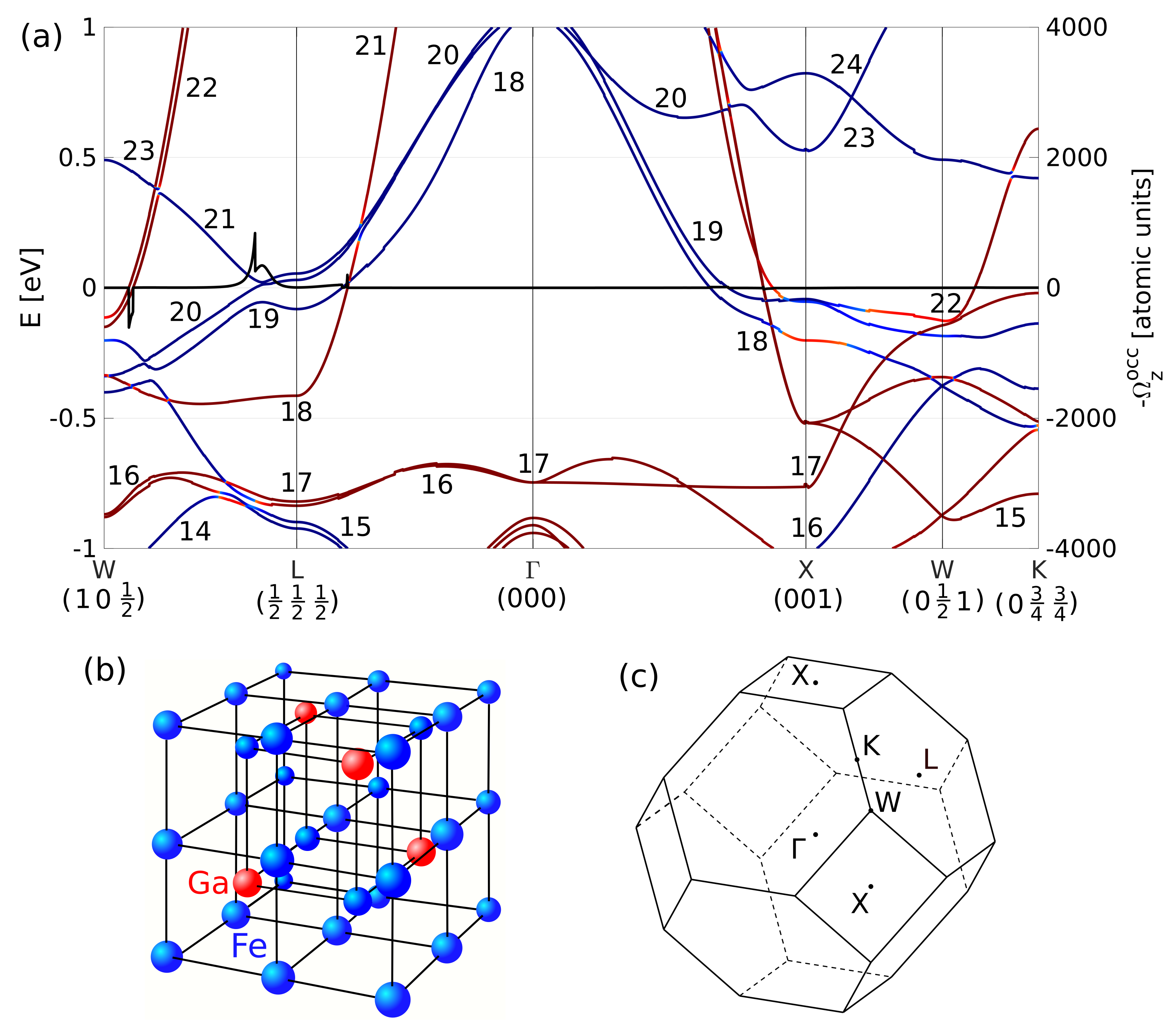}
\end{center}
\caption{(a) Band structure of DO$_3$ ordered Fe$_3$Ga crystal. Red (blue) color corresponds to spin down (up). Bands are indexed by their increasing energy eigenvalue starting with 3$d$ Fe bands. SOI is included. The black curve corresponds to the $z$-component of the Berry curvature over occupied states $\Omega_z^\mathrm{occ}=\Sigma_n f_n \Omega_z^n$. (b) Crystal structure of Fe$_3$Ga. (c) Brillouin zone of fcc crystal.
}
\label{fig:bsfull}
\end{figure}

The electronic structure of DO$_3$ ordered Fe$_3$Ga bulk crystal (Fig.~\ref{fig:bsfull}(a)) was calculated by the WIEN2k code~\cite{Wien2k,Draxl2006} with 27,000 $k$-points in the full Brillouin zone and the generalized gradient approximation~\cite{Perdew96} as the exchange-correlation potential. The lattice constant was set to 5.80 \AA~\cite{Matyunina2018}. The calculation was performed with the spin-orbit interaction and with magnetization in the $z$-direction. The product of the smallest atomic sphere and the largest reciprocal space vector was set to $R_\mathrm{MT}K_\mathrm{max}=7$ with the maximum value of the partial waves inside the spheres $l_\mathrm{max}=10$. States up to 3$s$ are treated as core states. Bands are labeled by an increasing energy eigenvalue starting with 3$d$ Fe states. The closest Ga states are located approximately 7\,eV below the Fermi level.

The Berry curvature is calculated by the Kubo-like formula:
\begin{equation}
\Omega_{\mu\nu}^n(\mathbf{k})=-\frac{\hbar^2}{m^2}\sum_{n'\neq n} \frac{2 \mathrm{Im}\bra{\psi_{n\mathbf{k}}}p_{\mu}\ket{\psi_{n'\mathbf{k}}}\bra{\psi_{n'\mathbf{k}}}p_{\nu}\ket{\psi_{n\mathbf{k}}}}{(E_n-E_{n'})^2},
\end{equation}
where $E_n$ is the band energy, $\psi_{n\mathbf{k}}$ are Bloch wave functions, and $p_\mu$ are momentum operators. The Berry curvature $\Omega_{\mu\nu}$, which is an antisymmetric tensor that in three dimensions can be expressed in a pseudovector form via $\Omega_{\mu\nu}=\epsilon_{\mu\nu\xi}\Omega_\xi$. The calculation of the anomalous Hall conductivity (AHC) is carried out by integration of the Berry curvature over the Brillouin zone:
\begin{equation}
\sigma^{\mathrm{AHE}}_{xy}=-\frac{e^2}{\hbar}\frac{1}{(2\pi)^3}\sum_n\int_\mathrm{BZ} f_n(\mathbf{k})\Omega_z^n(\mathbf{k})\,\mathrm{d}\mathbf{k}
\label{eq:sigma}
\end{equation}
where $f$ is the Fermi-Dirac distribution function. This is nowadays a well-established procedure for evaluating the intrinsic part of the anomalous Hall effect~\cite{Yao2004,Nagaosa2010,Xiao2010,Ernst2019,Helman2021}.

The strength of the ANE is evaluated by the temperature dependent anomalous transverse thermoelectric coefficient:
\begin{equation}
\alpha_{xy}(T)=-\frac{1}{e}\int \mathrm{d}E\left(-\frac{\partial f}{\partial E}\right) \sigma^{\mathrm{AHE}}_{xy}(E)\frac{E}{T}
\label{eq:alpha}
\end{equation}
The anomalous Hall conductivity $\sigma^{\mathrm{AHE}}_{xy}(E)$ becomes a function of energy $E$ that represents the threshold for which the electron states are treated as occupied. AHC is required to be calculated in a narrow energy interval in the vicinity of the Fermi level, therefore even minor changes in the form of deformation and doping can modify the value of ANE significantly.

%\begin{figure}
%\begin{center}
%\includegraphics[width=0.5\textwidth]{figs/struct.png}
%\end{center}
%\caption{Crystal structure of Fe$_3$Ga.
%}
%\label{fig:crystal}
%\end{figure}

\subsection{Deformation}

\begin{figure}
\begin{center}
\includegraphics[width=0.5\textwidth]{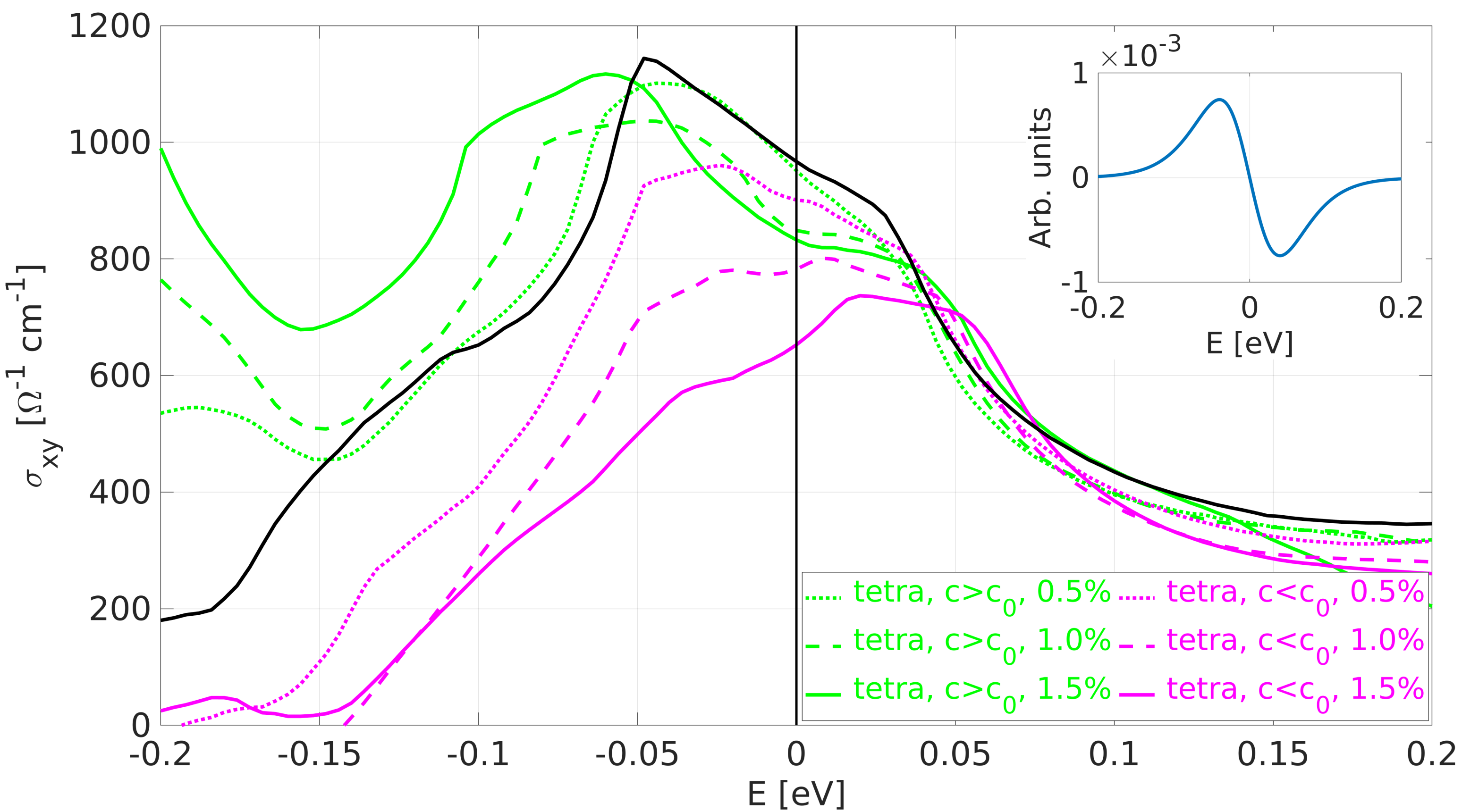}
\end{center}
\caption{AHC spectrum for deformed Fe$_3$Ga crystal structure. Black line represents spectrum of undeformed bulk AHC. In the inset, the weighting function for evaluation of $\alpha$ is shown.
}
\label{fig:deformation}
\end{figure}

Two general cases of deformation were considered, namely compressive and tensile strain. With compressive strain, the in-plane lattice parameter is shortened by 0.5, 1.0, and 1.5\%, while preserving the total volume of the unit cell. As the out-of-plane lattice parameter gets larger, this case is denoted by tetra, $c>c_0$.

In the tensile strain case, the in-plane lattice parameter is extended by 0.5, 1.0, and 1.5\%, denoted by tetra, $c<c_0$. In both cases, the crystal structure changes from cubic to tetragonal.

The results are shown in Fig.~\ref{fig:deformation} and the values of the thermoelectric coefficient $\alpha_\mathrm{max}$ (defined as the maximal/minimal value of $\alpha_{xy}(T)$) and $\alpha(300\,K)$ are summarized in Tab.~\ref{tab:alphamax}. In both cases, deformation leads to smoothening of the spectrum. However, while the tensile strain tends to cancel out the effect, the compression leads to a significant increase of ANE. Therefore, in thin film applications substrates with smaller lattice parameters should be considered.

In the inset of Fig.~\ref{fig:deformation} the evaluation at 300\,K of the function entering the convolution (Eq.~\ref{eq:alpha}) is depicted. This function weights the value of the anomalous Hall conductivity $\sigma^{\mathrm{AHE}}_{xy}(E)$ in the vicinity of the Fermi level to obtain the value of $\alpha$. For correct numerical evaluation $\sigma^{\mathrm{AHE}}_{xy}(E)$ needs to be calculated in the energy interval from $-0.2$\,eV to $0.2$\,eV.

{\renewcommand{\arraystretch}{1.2}
\begin{table}
\caption{Thermoelectric coefficients $\alpha_\mathrm{max}$ and $\alpha(300\,K)$ for deformed and doped Fe$_3$Ga}
\centering
\begin{tabular} {c c c}
\hline
crystal structure & $\alpha_\mathrm{max}$ [AK$^{-1}$m$^{-1}$] & $\alpha(300\,K)$ [AK$^{-1}$m$^{-1}$]  \\
\hline
cubic Fe$_3$Ga & 1.7 & 1.7 \\
\textcolor{green}{tetra, $c>c_0$, 0.5\%} & 2.1 & 2.1 \\
\textcolor{green}{tetra, $c>c_0$, 1.0\%} & 2.1 & 2.0 \\
\textcolor{green}{tetra, $c>c_0$, 1.5\%} & 2.4 & 2.2 \\
\textcolor{magenta}{tetra, $c<c_0, 0.5\%$} & 0.9 & 0.8 \\
\textcolor{magenta}{tetra, $c<c_0, 1.0\%$} & -0.6 & -0.1 \\
\textcolor{magenta}{tetra, $c<c_0, 1.5\%$} & -0.9 & -0.9 \\
\textcolor{red}{Fe$_3$Ga$_{0.95}$Ge$_{0.05}$} & 2.5 & 2.5 \\
\textcolor{red}{Fe$_3$Ga$_{0.9}$Ge$_{0.1}$} & 3.0 & 3.0 \\
\textcolor{red}{Fe$_3$Ga$_{0.8}$Ge$_{0.2}$} & 3.3 & 3.0 \\
\textcolor{blue}{Fe$_3$Ga$_{0.95}$Zn$_{0.05}$} & 0.8  & 0.6 \\
\textcolor{blue}{Fe$_3$Ga$_{0.9}$Zn$_{0.1}$} & -0.7 & -0.6 \\
\textcolor{blue}{Fe$_3$Ga$_{0.8}$Zn$_{0.2}$} & -2.6 & -2.6 \\
\hline
\end{tabular}
\label{tab:alphamax}
\end{table}
}

\subsection{Doping}

\begin{figure}
\begin{center}
\includegraphics[width=0.5\textwidth]{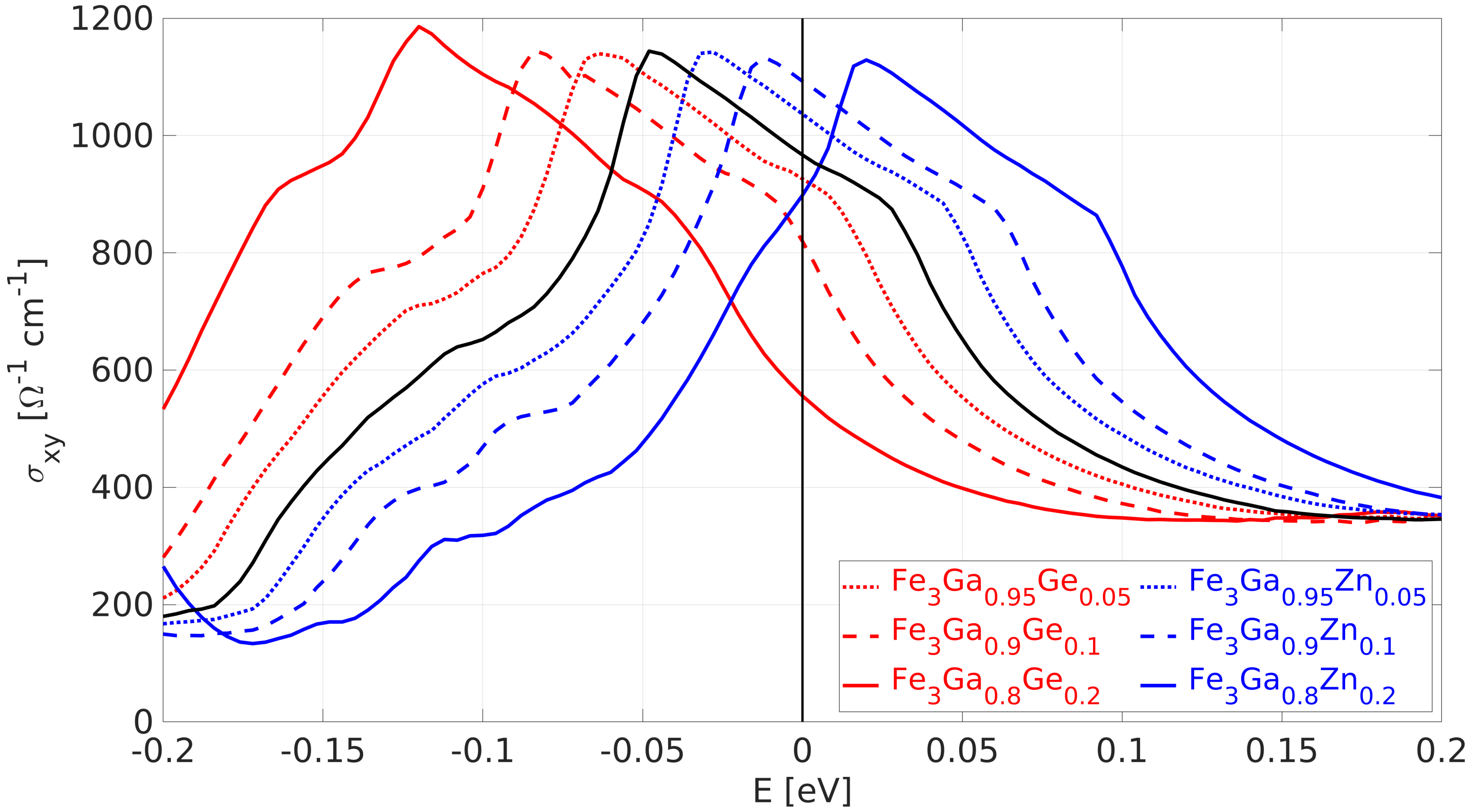}
\end{center}
\caption{AHC spectrum for doped Fe$_3$Ga crystal structure.
}
\label{fig:doping}
\end{figure}

The electronic structure of doped Fe$_3$Ga was modelled by the virtual crystal approximation. It is implemented by replacing Ga with a virtual element having non-integer atomic number, which simulates an effective increase or decrease of electron density. Six alloys were considered substituting 5, 10 and 20\% of Ga elements by either Ge or Zn.

From Fig.~\ref{fig:doping}, it is apparent that doping effectively moves the position of the Fermi energy as the Ga states are buried deeply below the Fermi level. Substituting Ga with heavier element (Ge) leads to a significant increase of the effect (see Tab.~\ref{tab:alphamax}) while lighter element (Zn) kills the effect and with higher concentration even changes its sign.

Generally, the value of $\alpha$ turns out to be very sensitive on the exact position of the Fermi level. Different ab-initio codes and applied potentials result in different values of the Fermi energy, making it a very difficult parameter to obtain from first principles alone. Sakai et al.~\cite{Sakai2020} reported $\alpha_\mathrm{max}=3.0$\,AK$^{-1}$m$^{-1}$ for Fe$_3$Ga. We arrive at the same value if we adjust the Fermi level to match theirs.

\section{Analysis of the ANE spectra}

\begin{figure}
\begin{center}
\includegraphics[width=0.5\textwidth]{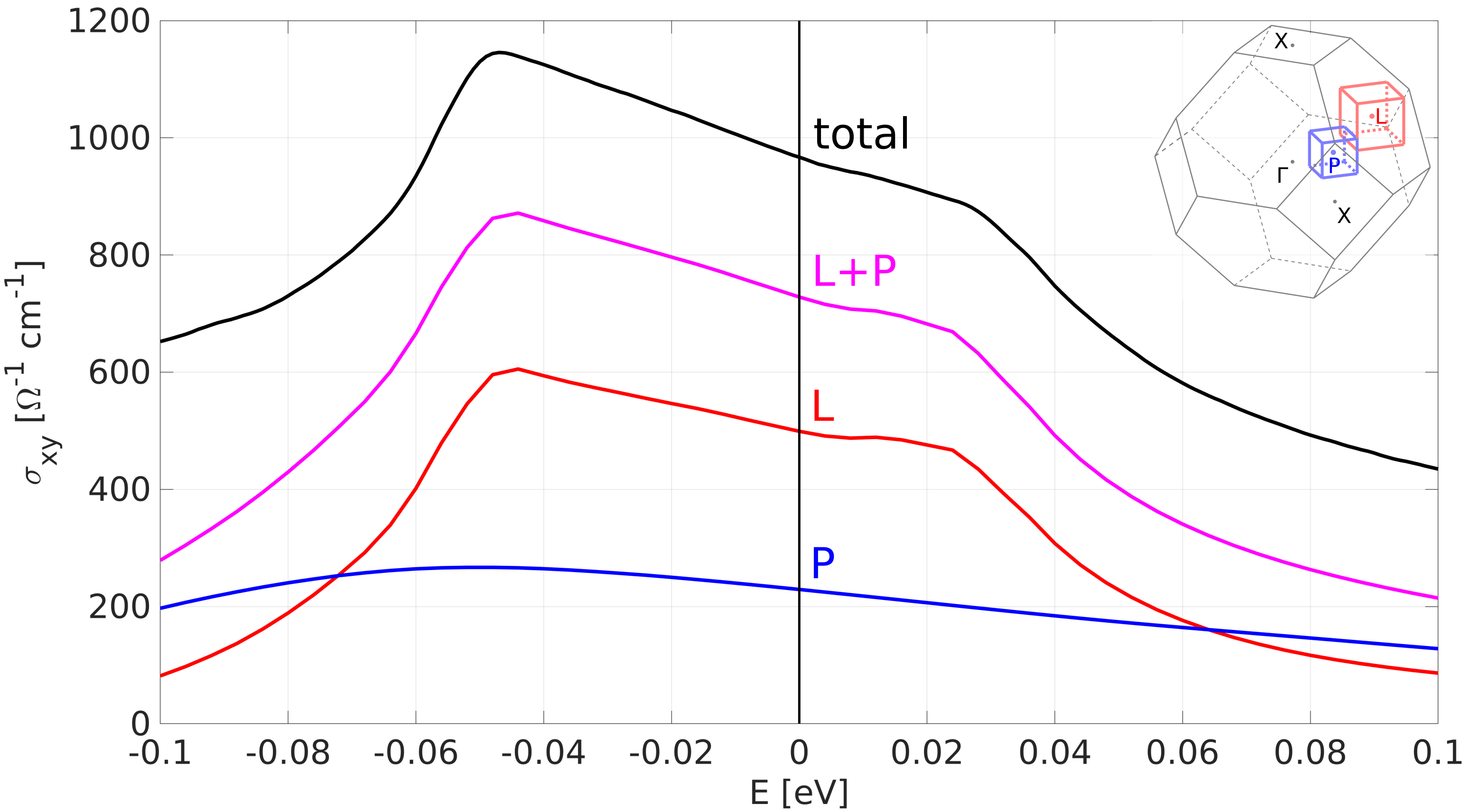}
\end{center}
\caption{Spectrum of the anomalous Hall conductivity in Fe$_3$Ga as a function of the Fermi energy position. The total value is represented by black line. Red, blue and magenta lines correspond to contributions to conductivity from L, P and L+P cubes, respectively. In the inset, the Brillouin zone is depicted with L and P cubes.
}
\label{fig:spectrum}
\end{figure}

In this section we analyze the ANE spectrum to identify the origin of the effect. On the standard $k$-path through the reciprocal space (see Fig.~\ref{fig:bsfull}(a)), there is only single major source of the Berry curvature in the vicinity of point L as identified by other authors~\cite{Sakai2020}. However, the inspection of the Brillouin zone reveals that there are two distinct contributions that cover the majority of AHC. Apart from the aforementioned L~point, strong source of the Berry curvature appears in general position around point P=[0.55,0.15,0.25], expressed in the relative Cartesian coordinates.

Both sources can be numerically separated and their partial spectra (integrated inside the respective cubes, including multiplicity) are shown in Fig.~\ref{fig:spectrum} along with the total anomalous conductivity obtained by integration of the Berry curvature over the entire Brillouin zone via Eq.~\ref{eq:sigma} in the energy range $-0.1$ to $0.1$\,eV. By modifying the value of the Fermi energy, only the positions of the Fermi surfaces change which is reflected by $f_n(\mathbf{k})$ entering the integral through $\mathbf{\Omega}^\mathrm{occ}=\Sigma_n f_n(\mathbf{k}) \mathbf{\Omega}^n(\mathbf{k})$. The Berry curvature vector field of individual bands $\mathbf{\Omega}^n(\mathbf{k})$ is not affected by the modification of electron occupancy.

%In the inset of Fig.~\ref{fig:spectrum} the anomalous transverse thermoelectric coefficient $\alpha_{xy}$ is presented being proportional to the energy derivative via the Mott relation:
%\begin{equation}
%\alpha_{xy}/T=-\frac{\pi^2k_{\mathrm{B}}^2}{3e}\left(\frac{\partial \sigma_{xy}}{\partial E}\right)_{E=E_\mathrm{F}}
%\end{equation}
Together, both partial spectra cover over 70\% of the total anomalous conductivity throughout the entire energy range and are separated from the total by a mere constant.%, providing identical values of $\alpha_{xy}$. To obtain large $\alpha_{xy}$, rapid change of the anomalous conductivity as the Fermi surfaces move through the reciprocal space is required.

In following, we shall analyze both contributions separately, identify the contributing Berry curvature flows and inspect the evolution of the Fermi surface with the Fermi level position.

\subsection{Contribution originating in cube P}

\begin{figure}
\begin{center}
\includegraphics[width=0.5\textwidth]{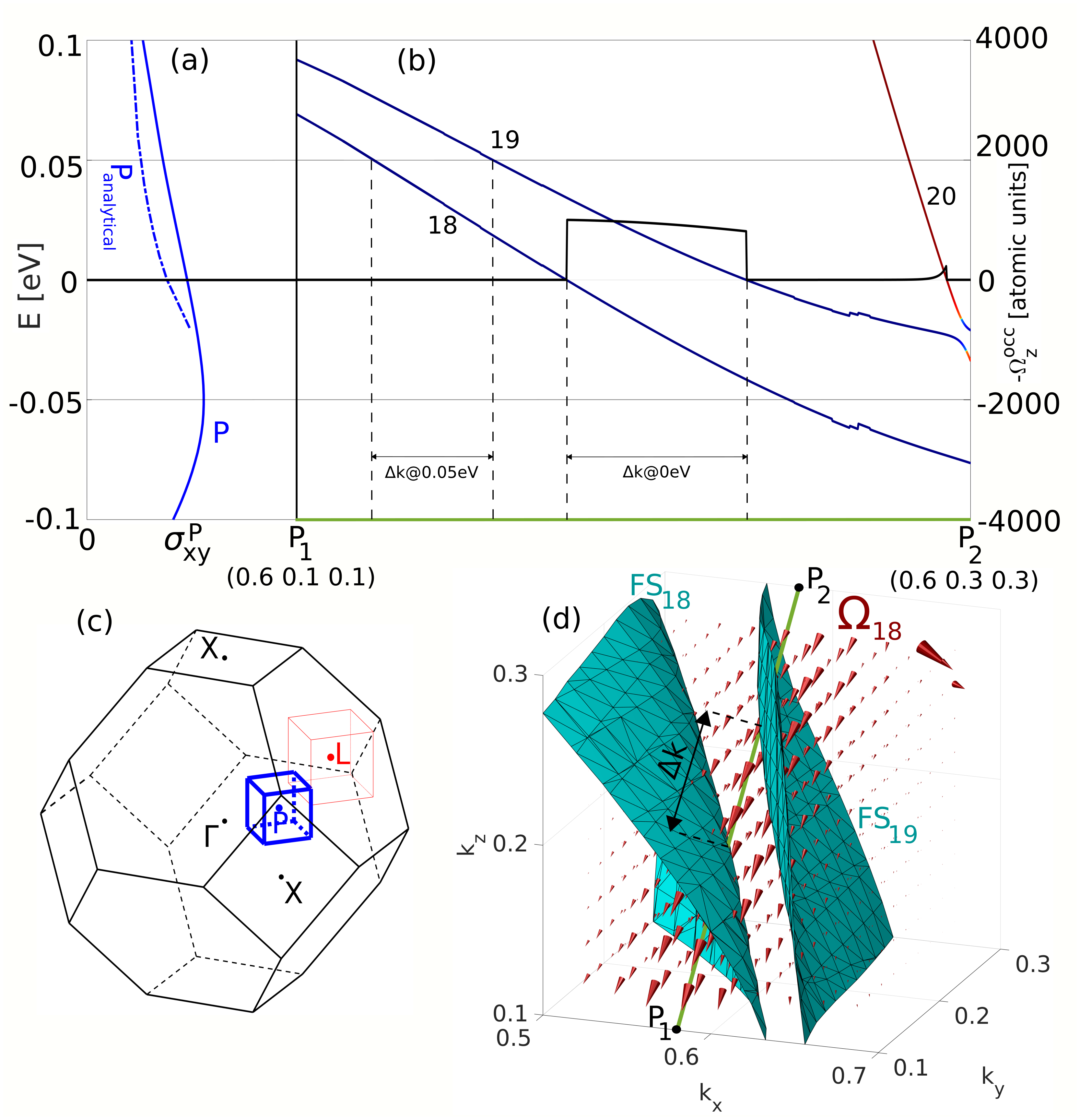}
\end{center}
\caption{a) AHC spectrum originated in cube P. The dash-dotted blue line corresponds to the analytical approximation of the P contribution (Eq.~\ref{eq:ahe}). b) Band structure in the direction of the Berry curvature flow with the $z$-component of the Berry curvature over occupied states $\Omega_z^\mathrm{occ}=\Sigma_n f_n \Omega_z^n$ that is approximately equal to $\Omega_z^{18}$ in the region between both Fermi surfaces. The reciprocal distance $\Delta k$ at two Fermi level positions is indicated. c) Brillouin zone with the position of cube P. d) Flow of the Berry curvature of band 18 in cube P. The flux of the flow equals 2$\pi$/3. Fermi surfaces for bands 18 and 19 at $E=0$\,eV are depicted, Berry curvature between the surfaces contributes to AHC. The green line represents the path along which the band structure in b) is depicted.
}
\label{fig:P2}
\end{figure}

First, the contribution from cube centered at point P is analyzed (see Fig.~\ref{fig:P2}(c)). The energy dependence of the anomalous conductivity in this region is shown in Fig.~\ref{fig:P2}(a). It repeats 16 times in the Brillouin zone due to symmetry and describes 21\% of the total anomalous conductivity. Upon close inspection, it is found that this region features only two bands, namely band 18 and 19. The Berry curvature of band 18 is depicted in Fig.~\ref{fig:P2}(d), while Berry curvature of band 19 flows in the opposite direction providing $\mathbf{\Omega}^{18}+\mathbf{\Omega}^{19}=0$ in this part of the BZ. It is of particular interest to plot the band structure in the direction of the flow (along the green P$_1$P$_2$ line) which is shown in Fig.~\ref{fig:P2}(b).

Both bands are split by the crystal field even without SOI and do not form nodal line. SOI only further increases the energy split. Since $\mathbf{\Omega}^{18}+\mathbf{\Omega}^{19}=0$ and $\mathbf{\Omega}^\mathrm{occ}=\Sigma_n f_n \mathbf{\Omega}^n$, it is clear that only the Berry curvature in the region between the Fermi surfaces of bands 18 and 19 contributes to the anomalous conductivity. Furthermore, the flow is localized along the green P$_1$P$_2$ line, hence the value of the conductivity can be approximated by an analytical formula~\cite{Stejskal2022}:
\begin{equation}
 \sigma^\mathrm{P}_{xy}=-\frac{e^2}{\hbar}\frac{1}{(2\pi)^3}\Phi^{18}\Delta k \cos{\varphi},\quad \cos{\varphi}=\hat{\Omega}\cdot\hat{z}.
\label{eq:ahe}
\end{equation}
The Berry curvature flux is $\Phi^{18}=\int \mathrm{d}\mathbf{S}\cdot \mathbf{\Omega}^{18}=2\pi/3$ highlighting the three-fold rotational symmetry present in the crystal. $\Delta k$ is the reciprocal distance between the respective Fermi surfaces in the direction of the Berry curvature flow. In the depicted case $\Delta k=0.0417$\,bohr$^{-1}$ and $\cos{\varphi}=-1/\sqrt{2}$, as the Berry curvature flows in the $[0\overline{1}\overline{1}]$ direction. This provides contribution from single 1D flow $\sigma^\mathrm{P}_{xy}=11.5$ ($\Omega$cm)$^{-1}$. Due to symmetry, this contribution appears sixteen times in the BZ with the total strength of 184 ($\Omega$cm)$^{-1}$ representing about 80\% of the numerically obtained value in cube P as the Fermi surfaces are not exactly parallel when crossing the Berry curvature flow.% but diverge. In our case, $\Delta k$ corresponds to the shortest distance thus it underestimates the actual value.

It is apparent from the band structure in Fig.~\ref{fig:P2}(b) that the distance $\Delta k$ changes with varying position of the Fermi level which is one of the ways to obtain non-zero derivative and hence non-zero ANE. The results of the analytical formula for $\sigma^\mathrm{P}_{xy}$ are represented by dash-dotted blue line in Fig.~\ref{fig:P2}(a). They are not provided in the entire range as for energies below -0.02\,eV third band (20) emerges and acts as an additional source of Berry curvature.

%Inside cube P, the Mott relation can be approximated by $\alpha_{xy}=A\left(\frac{\partial \Delta k}{\partial E}\right)_{E=E_\mathrm{F}}$, where $A=\frac{k_{\mathrm{B}}^2Te\Phi^{18} \cos{\varphi}}{24\hbar\pi}$. This formula enables to estimate the strength of the effect directly by inspecting the band structure.

%\begin{figure*}
%\begin{center}
%\includegraphics[width=\textwidth]{figs/P_1.png}
%\end{center}
%\caption{pasova struktura pro prispevek LX, paralelne s tokem, cesta vyznacena ve 3D obrazku s Berryho tokem. Berryho krivost dotycneho pasu v detailu kostky LX, modre vyznacena cesta pro pasovou strukturu, cyan - Fermiho plochy pro relevantni pasy, v meziplosi Berryho krivost prispiva do AHE.
%The dash-dotted blue line corresponds to the analytical approximation of the P contribution.
%}
%\label{fig:P1}
%\end{figure*}

\subsection{Contribution originating in cube L}

\begin{figure}
\begin{center}
\includegraphics[width=0.5\textwidth]{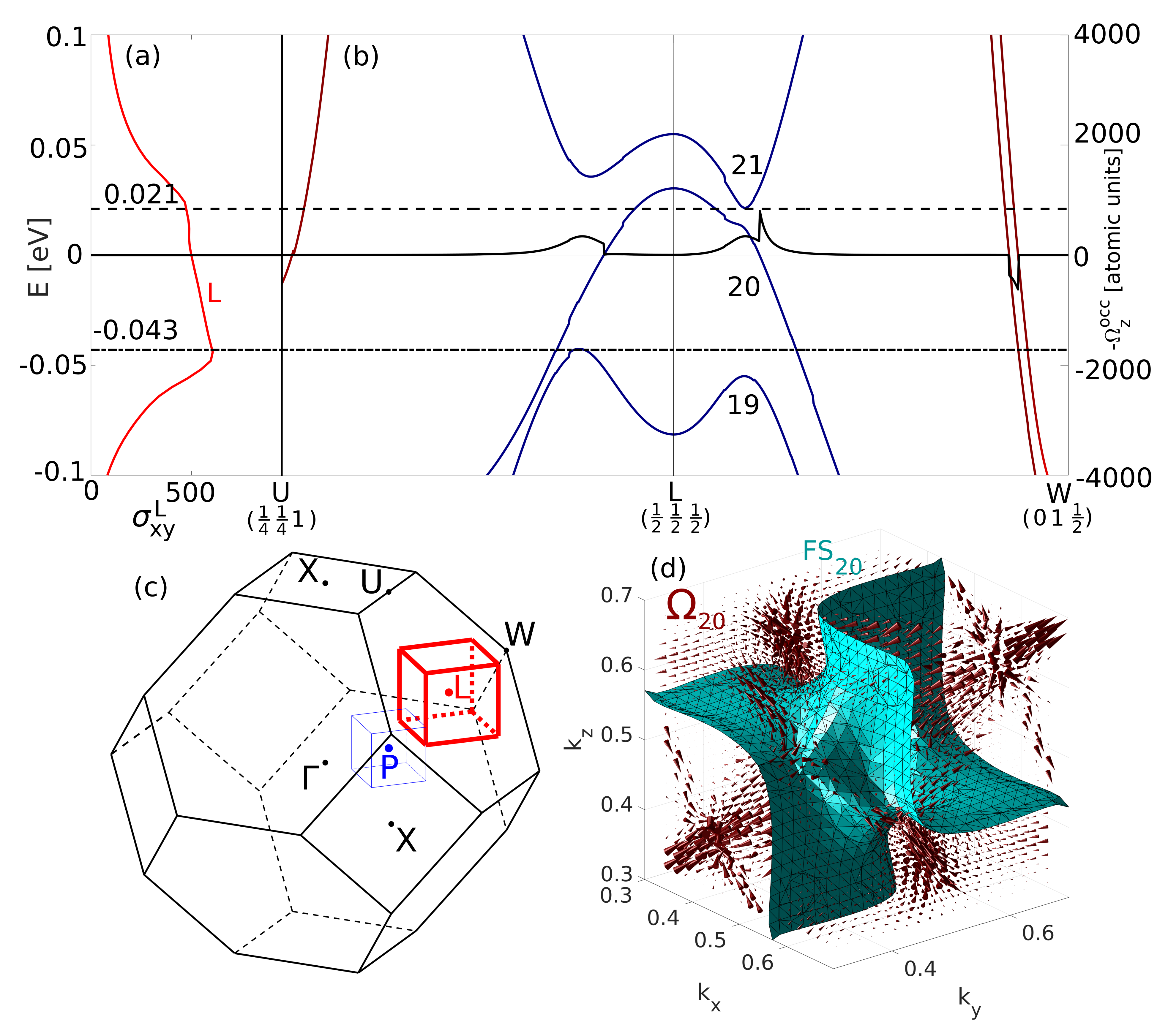}
\end{center}
\caption{a) AHC spectrum originated in cube L. b) Band structure along the ULW path. The dashed and dash-dotted lines at -0.043\,eV and 0.021\,eV, respectively, correspond to both turning points in the spectrum. c) Brillouin zone with the position of cube L. d) Flow of the Berry curvature and Fermi surface of band 20 in cube L. Berry curvature outside of the surface contributes to AHC.
}
\label{fig:L2}
\end{figure}

The second dominant source of the Berry curvature is located in the vicinity of the L~point (Fig.~\ref{fig:L2}(c)). The energy dependence of the anomalous conductivity in this region is shown in Fig.~\ref{fig:L2}(a). It repeats 4 times in the Brillouin zone due to symmetry and describes 52\% of the total anomalous conductivity. Even though there are no degeneracies present, the close proximity of the band triplet leads to strong hybridization and generates large Berry curvatures.

%Some authors refer to nodal lines as one-dimensional band degeneracies with the SOI present~\cite{Martinez2015}, while others use the term for degeneracies without SOI or even for bands closeby (approximately within 100\,meV) with no degeneracy at all. The terminology has not yet settled, however, it turns out to be unimportant as close energetic band proximity is sufficient for large anomalous effects, no degeneracies required.

The sum of the Berry curvatures of the band triplet $\mathbf{\Omega}^{19}+\mathbf{\Omega}^{20}+\mathbf{\Omega}^{21}$ equals zero in cube L. Therefore, if all three bands are either occupied or unoccupied, they do not contribute to the anomalous effects. The dashed and dash-dotted lines in the band structure Fig.~\ref{fig:L2}(b) mark the energies where the upper and lower band gets occupied and unoccupied, respectively, and correspond to both turning points in the spectrum Fig.~\ref{fig:L2}(a).

The anomalous effects inside energy region -0.043\,eV to 0.021\,eV (between dashed and dash-dotted lines) are governed solely by varying occupation of the middle band 20. The Berry curvature structure of this band is complicated and can be seen in Fig.~\ref{fig:L2}(d). The Berry curvature in the occupied region contributes to the anomalous conductivity. With increasing Fermi energy, the Fermi surface shrinks inwards enabling other parts of the Berry curvature flows to contribute. There are flows in many directions, some add to and some subtract from the total value. Thus, precise geometry and width of the individual flows matter and the value of AHC steadily decreases with increasing energy leading to non-zero contribution to ANE.

%\begin{figure*}
%\begin{center}
%\includegraphics[width=\textwidth]{figs/L_1.png}
%\end{center}
%\caption{pasova struktura ULW, vyznaceny priciny dvojice hrbu ve spektru. DO SUPPLEMENTU. Berryho krivost pro relevantni pas vcetne Fermiho plochy. S posunutim Fermiho energie se plocha posouva. Do AHE prispiva vse vne plochy.
%}
%\label{fig:L1}
%\end{figure*}

\subsection{Partial contributions under deformation}

\begin{figure}
\begin{center}
\includegraphics[width=0.5\textwidth]{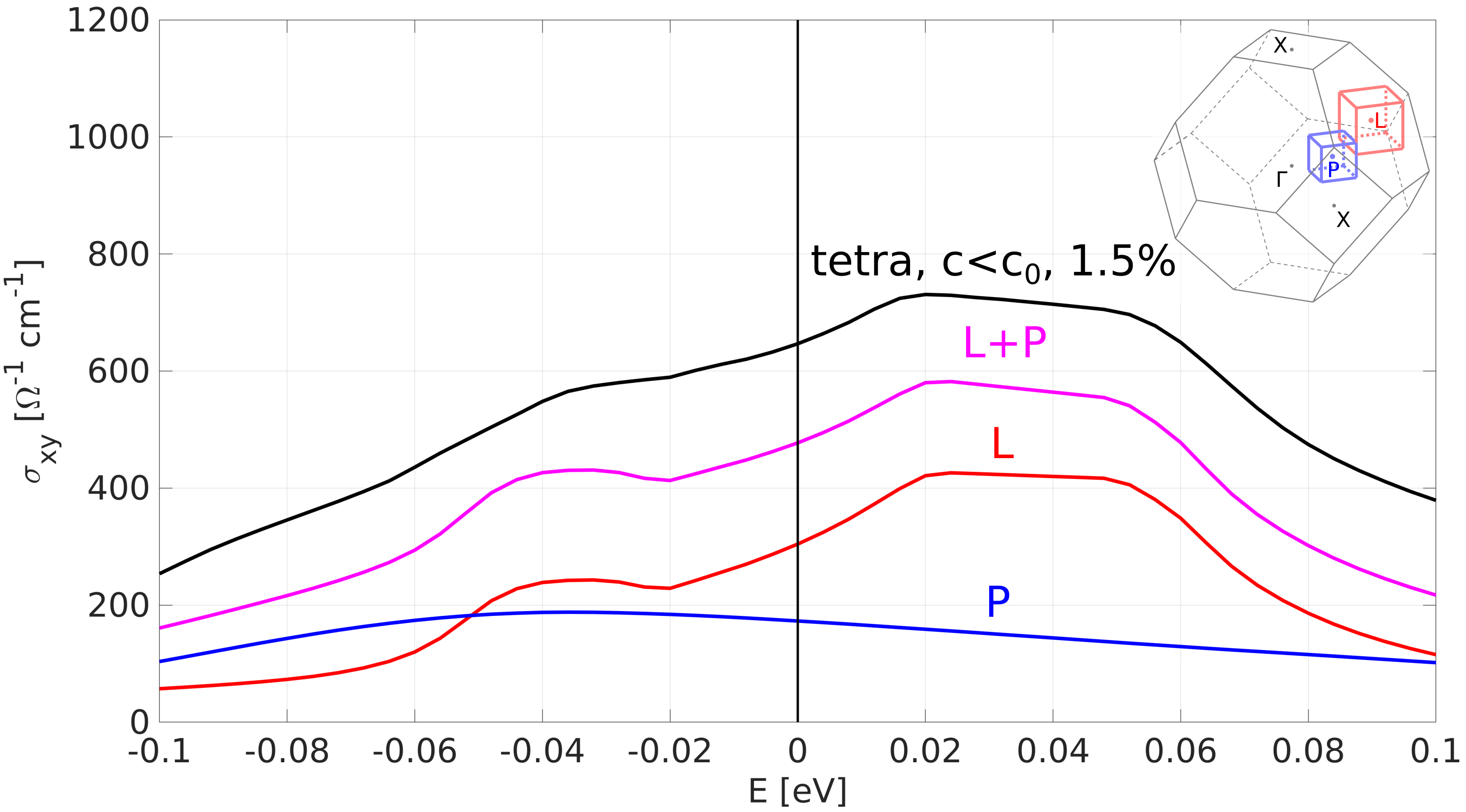}
\end{center}
\caption{AHC spectrum of stretched Fe3Ga with partial contributions.
}
\label{fig:tetra2}
\end{figure}

\begin{figure}
\begin{center}
\includegraphics[width=0.5\textwidth]{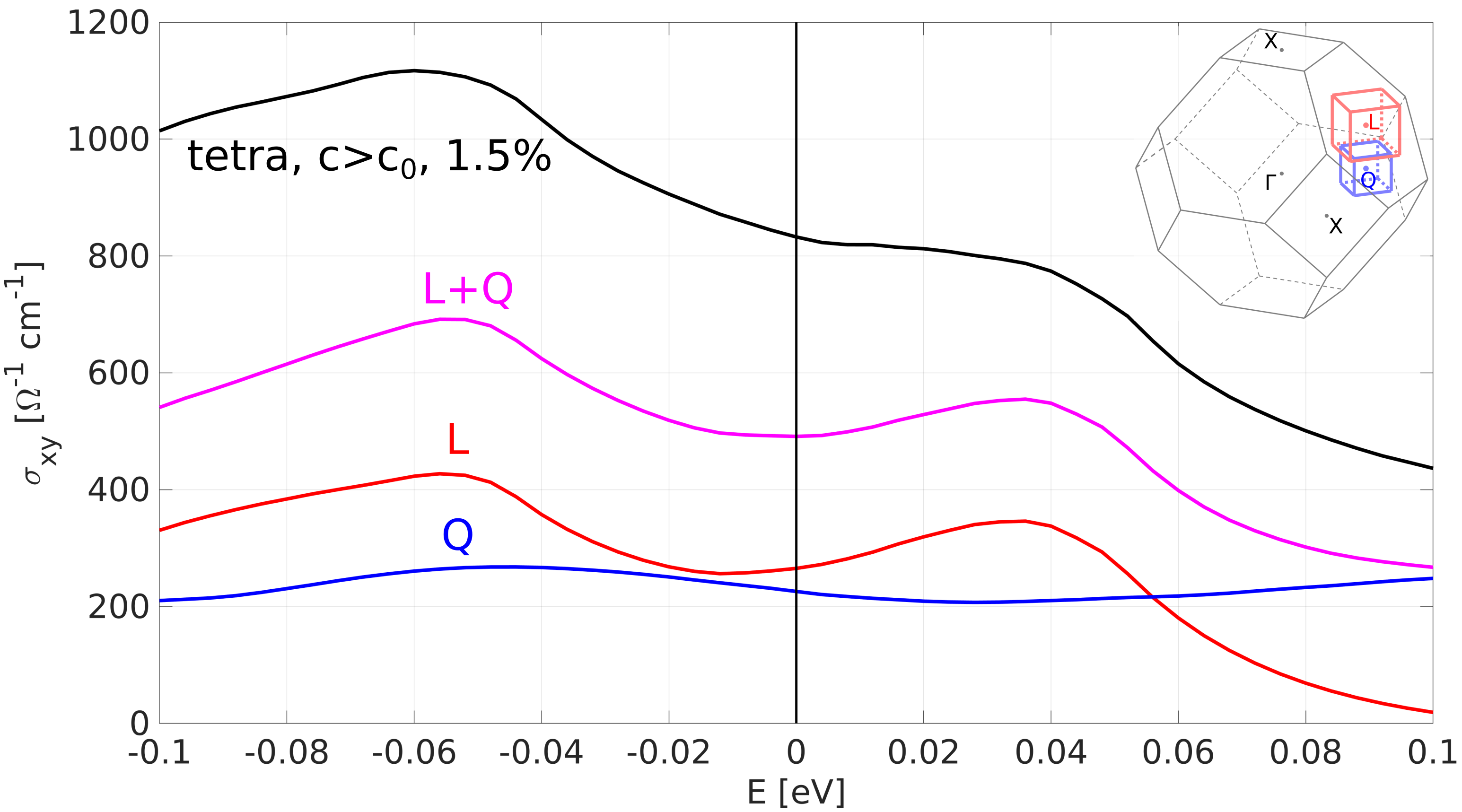}
\end{center}
\caption{AHC spectrum of compressed Fe3Ga with partial contributions.
}
\label{fig:tetra1}
\end{figure}

Under tensile (Fig.~\ref{fig:tetra2}) and compression (Fig.~\ref{fig:tetra1}) strain, the partial contributions analysis shows that the vicinity of point L dominates both AHC spectra and determines solely their overall shape.

In the compressed case, the P contribution vanishes (as the energy distance between respective bands increased) but is substituted by newly occurring Q=[0.5,0.5,0.15] contribution of similar strength (see the inset).

Overall character of the spectra is preserved under deformation, with both fundamental peaks approximately at the same energy position, suggesting small change in the electronic structure.
On the other hand, the strength of individual peaks decreases, suggesting change of spin-orbit splitting between respective bands leading to modifications of the Berry curvature vector field.
%\section{Deformation and doping}

%To investigate possible enhancements of ANE, we performed several calculations of deformed and doped Fe$_3$Ga. The resulting spectra are presented in Fig.~\ref{fig:modif} and the values of the thermoelectric coefficient $\alpha_\mathrm{max}$ (see Appendix~\ref{sec:compmeth}) are summarized in Tab.~\ref{tab:alphamax}.

%In the deformation case, the in-plane lattice parameter was shortened (extended) by 1.5\% denoted by tetra, $c>c_0$ (tetra, $c<c_0$) respectively. The total volume of the unit cell was preserved, but the crystal structure changed to tetragonal. In both cases, it smoothened the spectrum, however, for tetra, $c>c_0$, i.e. out-of-plane lattice parameter larger than the in-plane one, the value of $\alpha_\mathrm{max}$ increases.

%Two doped Fe$_3$Ga were calculated, namely Fe$_3$Ga$_{0.8}$Ge$_{0.2}$ and Fe$_3$Ga$_{0.8}$Zn$_{0.2}$, simulating an effective increase or decrease of electron density. From Fig.~\ref{fig:modif}, it is apparent that doping moves the position of the Fermi level, however, at the same time it slightly smoothens out the spectrum decreasing its derivative. As the Fermi energy shifts towards steeper parts of the spectrum, in both cases, we observed increase of $\alpha_\mathrm{max}$.

%\begin{figure}
%\begin{center}
%\includegraphics[width=0.5\textwidth]{figs/fig_DOP.png}
%\end{center}
%\caption{AHC spectrum for deformed and doped Fe$_3$Ga crystal structure.
%}
%\label{fig:modif}
%\end{figure}

%%% CONCLUSION %%%
\section{Conclusion}

We performed ab-initio calculations of several Fe$_3$Ga structures modified by deformation and doping. We showed that compressive strain should be sought for in applications as it leads to a significant increase of $\alpha_{xy}$. Doping modelled via the virtual crystal approximation effectively moves the position of the Fermi level and enables additional tuning of the $\alpha_{xy}$ value.

Furthermore, we analyzed the origin of the anomalous Nernst effect in DO$_3$ ordered Fe$_3$Ga and identified two distinct sources in the Brillouin zone. Note however, that the analysis presented in this work covers only the intrinsic part of the anomalous Nernst effect and omits scattering mechanisms.

\section*{Acknowledgements}

This work was supported by the Czech Science Foundation (GACR), Grant No. GA19-13310S and by the European Union project Matfun, Project No. CZ.02.1.01/0.0/0.0/15\_003/0000487.

%\appendix
%\section{Computational method}
%\label{sec:compmeth}

%The temperature dependence of the anomalous transverse thermoelectric coefficient $\alpha_{xy}$ was calculated by:
% $\alpha_\mathrm{max}$ corresponds to the maximum/minimum value of $\alpha_{xy}(T,0)$.

%\bibliographystyle{h-physrev.bst}
%\bibliography{Fe3Gapaper}

\end{document}